\documentclass[twocolumn,prl]{revtex4}

\newcommand{\figurewidth}{84mm}
\newcommand{\kk}{\mathbf{k}}

\usepackage{graphicx}
\usepackage{dcolumn}

\newcommand{\beq}{\begin{equation}}
\newcommand{\eeq}{\end{equation}}
\newcommand{\Tr}{\mbox{Tr}}
\newcommand{\leqs}{\stackrel{\scriptstyle<}{\scriptscriptstyle\sim}\:}

\begin{document}

\title[Short Title]{
   Lowering of the Kinetic Energy in Interacting Quantum Systems
}
 
\author{Burkhard Militzer and E. L. Pollock}
\affiliation{ Physics Department,\\
              Lawrence Livermore National Laboratory,\\
              University of California, Livermore, California 94550 }
 
\date{\today}

\begin{abstract}
    
Interactions never lower the ground state kinetic energy of a quantum
system. However, at nonzero temperature, where the system occupies a
thermal distribution of states, interactions can reduce the kinetic
energy below the noninteracting value. This can be demonstrated from a
first order weak coupling expansion. Simulations (both variational and
restricted path integral Monte Carlo) of the electron gas model and
dense hydrogen confirm this and show that in contrast to the ground
state case, at nonzero temperature the population of low momentum
states can be increased relative to the free Fermi distribution. This
effect is not seen in simulations of liquid $^3$He.
\end{abstract}
 
\maketitle

\section{Introduction}

It is a common assumption that the addition of interactions to a
noninteracting quantum system will broaden the momentum distribution
and increase the kinetic energy (proportional to the second moment of
the momentum distribution). An intuitive understanding of this follows
from perhaps the first example encountered in studying quantum
mechanics, the particle in a box, where the kinetic energy scales as
the box size $L^{-2}$.  More generally the kinetic energy, given by
the average curvature of the wavefunction, is expected to scale as the
``localization length''$^{-2}$.  Since interactions lead to a
nonuniform density, i.e. increased ``localization'', the kinetic
energy is expected to increase.

\begin{figure}
\includegraphics[angle=0,width=\figurewidth]{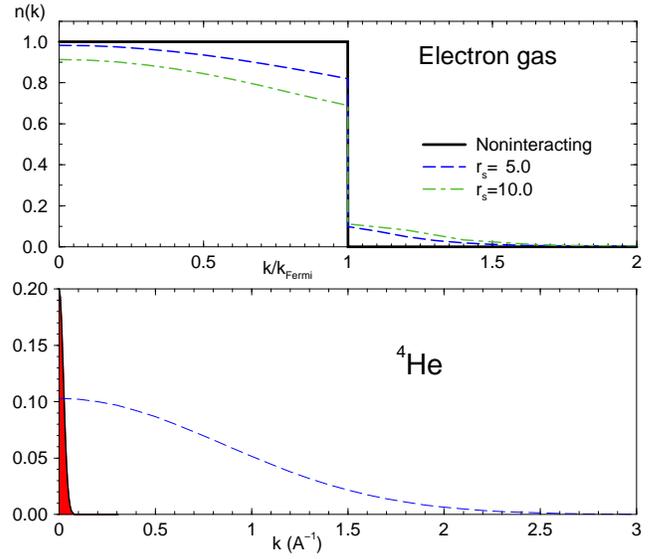}
\caption{ Top figure: Broadening of the ground state electron gas
            momentum distribution at $r_{s}=5$ (dashed line) and
            $r_{s}=10$ (dot-dashed line) caused by interactions.
            $r_{s}$ is the radius, in units of the Bohr radius, of a sphere 
            containing one electron and
            in metallic elements at atmospheric pressure ranges from roughly
            2 to 6. The unit discontinuity of the momentum distribution
            for the noninteracting system (solid line)
            at the Fermi surface is reduced by $28 \%$ and $42 \%$
            respectively. Corresponding increases in the kinetic energy
            over the noninteracting case are $30 \%$ and $60 \%$.
            Bottom figure: For $^{4}$He at $T=0$ and $P\approx 0$ roughly 92\% of 
            the zero momentum condensate (black vertical bar along y axis) in
            the noninteracting Bose system is promoted to
            higher momentum states (distribution given by dashed line)
            leading to a kinetic energy increase of over 14~K per particle.
            }
\label{FIG1}
\end{figure}

For condensed matter scientists, two well ingrained many body examples
of this, illustrated in Fig.~\ref{FIG1}, are the broadening of the
free Fermi ground state momentum distribution due to electron
repulsion and the depletion of the zero momentum condensate in a
strongly interacting Bose system such as $^4$He~\cite{penrose}.  For
the homogeneous electron gas, Fig.~\ref{FIG1}a shows the promotion of
low momentum states to higher momentum and the reduction in the step
discontinuity at the Fermi surface~\cite{ortiz}. Similarly
Fig.~\ref{FIG1}b shows that the 100\% condensation into the zero
momentum ground state of a noninteracting Bose system is reduced to
roughly 8\% in $^{4}$He with the rest going into an almost Gaussian
distribution~\cite{pollock}.

The proof of this assumption for systems at zero temperature is an
immediate consequence of the ground state variational principle
applied to the free particle Hamiltonian, $H_{0}$, which states that
$K_{0}=\left<\Psi_0|H_0|\Psi_0\right>$ is minimized by $\Psi_0$, the true
ground state wavefunction of $H_0$. Using any other wavefunction,
such as the ground state, $\Psi_G$, for an interacting system,
$H=H_0+U$, leads to
\beq
K=\left<\Psi_G|H_{0}|\Psi_G\right>\;\geq K_{0} 
\;,
\label{eq1}
\eeq
demonstrating that the kinetic energy, $K$, of the interacting system
is never lower than the free particle kinetic energy, $K_{0}$.

This argument fails however, when generalized to nonzero temperature,
$T$, where the system occupies a thermal distribution of energy
eigenstates. From the Gibbs variational principle, the free energy
functional,
\beq
F[\rho]=\Tr[H_{0}\rho]+k_{B}T\;\Tr[\rho\ln\rho]     \;,
\label{eq2}
\eeq
takes its minimum value $F[\rho_{0}]=K_{0}-T S_{0} $
for the equilibrium density operator
\mbox{$\rho_{0} =e^{-\beta H_{0}}/\Tr [e^{-\beta H_{0}}] $} 
where $\beta=1/k_{B}T$ and $k_{B}$ is Boltzmann's constant.
Any other normalized density operator, such as that for the
interacting system, \mbox{ $\rho=e^{-\beta H}/\Tr [e^{-\beta H}] $},
gives a higher value,
\beq
F[\rho]=K-T S \;\; \geq \;\; F[\rho_{0}]=K_{0}-T S_{0}  \label{eq3}\;.
\eeq
The ground state inequality, Eq.~\ref{eq1},  now generalizes to,
\beq
K \geq K_{0} + T (S - S_{0}) \;.
                                         \label{eq4}
\eeq

Since interactions often increase order, the entropy $S$ of the
interacting system can be less than that of the noninteracting,
$S_{0}$, and $S-S_{0}$ may be negative. This allows Eq.~\ref{eq4} to
be satisfied while the ground state inequality, $K \geq K_{0}$, is
not. In general, at nonzero temperature nothing forbids interactions
from lowering the kinetic energy. 

      This kinetic energy lowering would only be observed in an intermediate 
range of temperatures since in the high temperature, classical limit,
interactions only effect equilibration rates, not the final Maxwellian 
momentum distribution.

\section{Weak Coupling Limit}

Not only is this lowering theoretically possible, it can be shown to
occur for a variety of weakly interacting physical systems using
lowest order thermodynamic perturbation theory. The change in the
Helmholtz free energy, $F$, to first order in the interaction
potential is
\beq
F = F_{0} +\left<U\right>_0 \;,                             \label{eq5}
\eeq
where $\left<U\right>_{0}$ is the potential averaged over the free particle
configurations. Since the total energy is given by, $E=\partial(\beta
F)/\partial\beta$, the first order change in the kinetic energy from
Eq.~\ref{eq5} is
\beq
K_1  = \beta \frac{\partial\left<U\right>_0}{\partial \beta} \;\;.    
\eeq
For a single component system of $N$ particles with pair interaction,
$V(r)$, the first order energy change reads,
\beq
\left<U\right>_{0}=N \frac{n}{2} \int g_{0}(r) \, V(r) \, dr^{3}    \label{eq7}
\eeq
where $n$ is number density. The first order kinetic energy change
\beq
K_{1} = -T N \frac{n}{2} \int 
\frac{\partial g_{0}(r)}{\partial T} \, V(r) \, dr^{3} 
                                               \label{eq8}
\eeq
depends on the temperature dependence of the free particle radial distribution
function, $g_{0}(r)$.

\begin{figure}
\includegraphics[angle=0,width=\figurewidth]{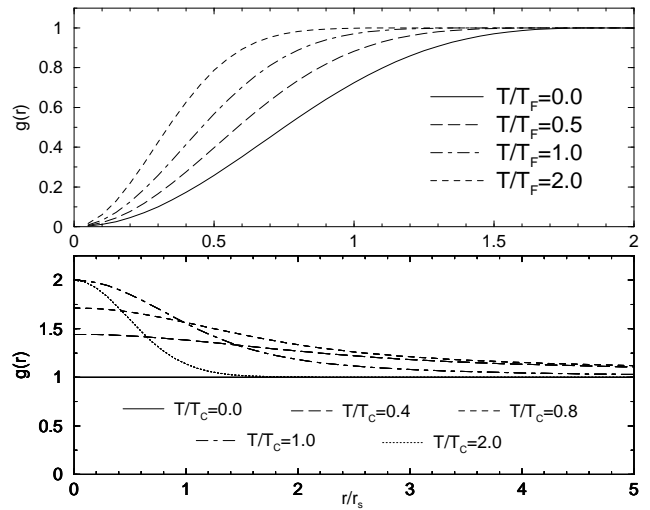}
\caption{ 
          Behavior of free fermion (top figure) and free boson (lower
          figure) radial distribution functions for indicated
          temperatures measured in terms of the Fermi temperature
          $T_{F}$ or condensation temperature $T_{c}$ respectively. }
\label{FIG2}
\end{figure}

As shown in Fig.~\ref{FIG2}, for both fermions and bosons $\partial
g_{0}(r)/\partial T$ can be positive at all $r$ leading to a kinetic
energy decrease for repulsive interactions although for quite
different reasons in the two cases. For fermions the increase of
$g_{0}(r)$ with temperature is due to the filling in of the ``Fermi
hole'', the region near the origin where due to the Pauli principle
like spin particles are excluded.  The size of this region is roughly
the de Broglie thermal wavelength which is proportional to $1/\sqrt{T}$.

For noninteracting bosons at $T=0$, all particles are in the zero
momentum condensate and $g_{0}(r)=1$. As the temperature increases
particles are promoted from the condensate and $g_{0}(r)$ increases at
all $r$ for $T/T_{c}\leq (2/5)^{2/3}$, where $T_{c}$ is the Bose
condensation temperature, and thereafter at small $r$, reaching
$g_{0}(0)=2$ at $T=T_{c}$~\cite{landau,baym}. Since $T\partial
g_{0}(r)/\partial T$ vanishes at $T=0$, the ground state inequality,
Eq.~\ref{eq1}, is not violated.

\begin{figure}
\includegraphics[angle=0,width=\figurewidth]{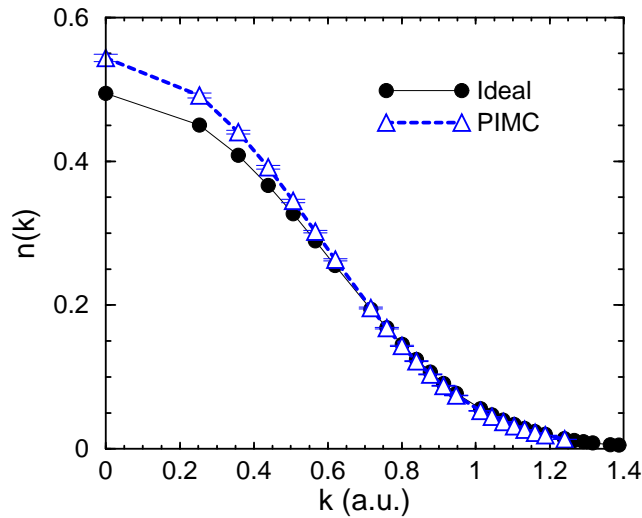}
\caption{ 
         Example of momentum distribution narrowing in the fully polarized electron gas
         at $r_{s}=4$ and $T=T_{\rm Fermi}$ from simulations with 57 particles.}
\label{FIG3}
\end{figure}

The mechanism for the narrowing of the electron gas momentum
distribution can be seen from the first order shift in the free
electron energy levels~\cite{ashcroft},
\beq
\Delta \epsilon(\kk) = - \frac{4 \pi e^2}{\Omega} \sum_{\kk' \ne \kk} \frac{n_0(\kk)}{|\kk-\kk'|^2}
\;\;.
\eeq
The decrease of $n_0(\kk)$ with k leads to a larger lowering of the
energy levels at smaller k and thus to a population redistribution
and momentum distribution narrowing as
demonstrated in Fig.~\ref{FIG3}.

\section{Simulation results}

For the electron gas, this weak coupling expansion applies only at
very high densities, $r_{s} \leqs 0.5$, and is of little use for
realistic condensed matter densities. It is necessary to use more
powerful quantum many body simulation methods. Two such methods, path
integral Monte Carlo (PIMC)~\cite{ceperley,ceperley2} and a variational trial
density matrix method (VDM)~\cite{pollock2}, have been used here to
search for narrowing of the momentum distribution in a variety of
fermion systems.

Liquid $^3$He is a strongly coupled system. Due to its steeply
repulsive short ranged interactions the weak coupling arguments of the
preceding paragraphs do not apply.  Restricted PIMC calculations using
free particle nodes for $^3$He at number density $n=0.01636 {\rm
\AA}^{-3}$ and temperatures from $1$ to $40$ K show the kinetic energy
to be more than $5$ K above the noninteracting value. Narrowing of the
momentum distribution is not found for this system.

For the electron gas model both restricted PIMC and VDM calculations
find momentum distribution narrowing. The effect is small, with the
relative decrease of the kinetic energy from its ideal value less than
a few percent~\cite{pokrant,kraeft}, but well within the accuracy of
the two methods. The computation of momentum distribution for fermions
with PIMC (see Fig.~\ref{FIG3}) required extending the restricted path
integral technique to simulations with open
paths~\cite{openpaths,openpaths2}.
\begin{figure}
\includegraphics[angle=0,width=\figurewidth]{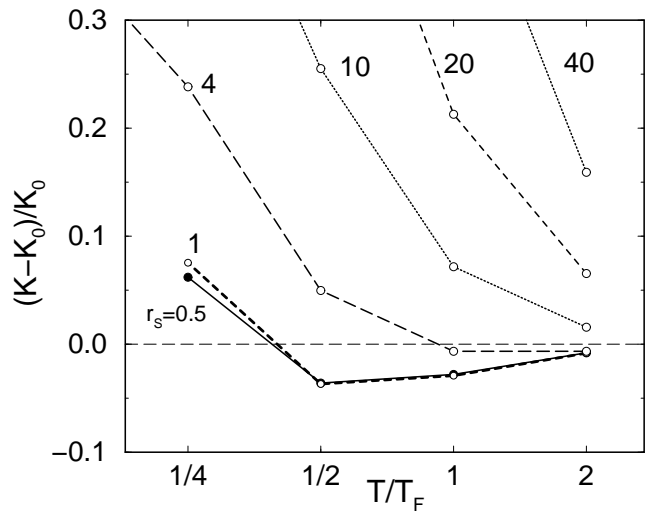}
\caption{ 
         Relative excess kinetic energy vs temperature for the
         unpolarized electron gas  at indicated
         $r_s$ values as calculated from PIMC simulations with 66
         particles in the periodic cell. } 
\label{FIG4}
\end{figure}

\begin{figure}
\includegraphics[angle=0,width=\figurewidth]{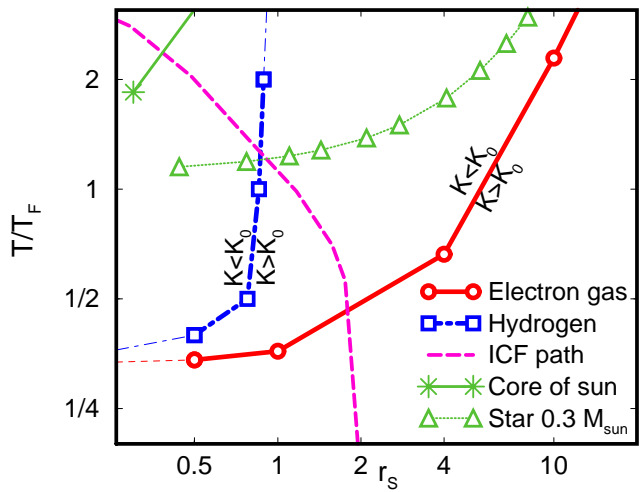}
\caption{
         Proposed region for momentum narrowing ($K<K_0$) in the
         unpolarized electron gas (area above the solid line) and in
         hydrogen plasma (area above the dot dashed line) as derived
         from PIMC simulations. The dashed line indicates the
         compression path of inertial confinement
         fusion~\cite{lindl}. The conditions near the core of the sun
         and a low mass star (0.3 solar masses) are also
         indicated~\cite{Saumon}. }
\label{FIG5}
\end{figure}

Fig.~\ref{FIG4} shows the difference in kinetic energy of the
homogeneous electron gas for different densities and temperatures
indicated by the degeneracy parameter $T/T_{\rm Fermi}$. For densities
corresponding to $r_s \leqs 4$, a lowering of the kinetic
energy with respect to the noninteracting value is found. The magnitude of the
lowering increases with density (smaller $r_s$). The statistical 
uncertainty of these results was estimated at $r_s=1$ and $T=T_{\rm Fermi}$
by a study of finite size effects (using 14, 38, 66, and 114 particles in
the periodic simulation cell) and path discretization errors
(using 16, 32, and 64 time steps) which gave 
$(K-K_0)/K_0 = 0.023 \pm 0.002$.  The region of the effect, as predicted 
by PIMC simulations, is shown in the high temperature and density 
phase diagram in Fig.~\ref{FIG5}.

Simulations of dense hydrogen plasma~\cite{pierleoni} have also
indicated kinetic energy lowering, which we confirmed and extended. We
found a maximal lowering of $(K-K_0)/K_0 = 0.007 \pm 0.001$ for
$r_s=0.5$ and $T=T_{\rm Fermi}$. The magnitude is reduced compared to
the electron gas because Coulomb interactions of electrons and protons
counteract the entropic lowering. This also results in a reduction of
the parameter region in the temperature and density plane for which
the effect can be observed. However, in the limit of high density, the
boundaries for the electron gas model and for hydrogen converge
because interactions become weaker in the high density
limit. Fig.~\ref{FIG5} shows that the lowering region of hydrogen
includes conditions near the core of the sun or other low mass
stars~\cite{Saumon} as well as on the compression path of inertial
confinement fusion~\cite{lindl}, indicating that hydrogen is a
potential candidate for experimental verification of this effect.

\section{Conclusions}

Analytical and numerical arguments show that interactions can reduce
the kinetic energy of a quantum system below the corresponding
noninteracting value. The associated narrowing in the momentum
distribution can potentially be verified experimentally. Prime
examples are the Bose condensates in magnetic traps, in which the
momentum distribution can be inferred by measuring the expansion rates
after the magnetic field has been switched off. Sufficiently accurate
measurements of the momentum distribution could be used to provide
information about the nature of the interactions of the trapped atoms.

Lowering of the kinetic energy was demonstrated in the homogeneous
electron gas model, one of the fundamental models in condensed matter
physics and a well studied example of a one-component
plasma. Corrections to the noninteracting kinetic energy are always
positive at zero temperature and vanish in the high temperature limit,
which makes it counterintuitive to expect lowering at intermediate
temperature. Research on the one-component plasma has focused more on
the internal energy, pressure and correlation energy. However, the
kinetic energy is relevant of our understanding of quantum systems and
can be determined experimentally by measuring the momentum
distribution.

The effect, previously indicated but left uninterpreted in dense
hydrogen~\cite{pierleoni}, is reconfirmed here.  The arguments
presented above suggest that the kinetic energy reduction should be
present in weakly coupled systems with repulsive pair
interactions. Simulation evidence for this effect was presented here
only for Coulombic systems.  More work is necessary to understand how
these arguments apply to other systems and where the entropic
reordering in the thermal population of states leading to a lower
kinetic energy can best be observed experimentally.

\section{Acknowledgements}

We thank Bernie Alder, Erich Mueller, and Forrest Rogers for their
comments on this work and Hugh DeWitt for triggering interest in
it. B.M. thanks David Ceperley for advice on the PIMC computation of
the momentum distribution. This work was performed under the auspices
of the U.S. Department of Energy by University of California Lawrence
Livermore National Laboratory under contract No. W-7405-Eng-48.

\end{document}